# Raman spectroscopy and electrical properties of InAs nanowires with local oxidation enabled by substrate micro-trenches and laser irradiation

R.Tanta[1], M.H.Madsen[3], Z.Liao[2], P. Krogstrup[1], T.Vosch[2], J.Nygård[1], T.S.Jespersen[1a)]

[1] *Center for Quantum Devices and Nano Science Center, Niels Bohr Institute, University of Copenhagen, Copenhagen, 2100, Denmark*

[2] *Nano-Science Center, Department of Chemistry, University of Copenhagen, Copenhagen, 2100, Denmark*

[3] *Danish Fundamental Metrology, Matematiktorvet 307, Kgs. Lyngby, 2800, Denmark*

The thermal gradient along indium-arsenide nanowires was engineered by a combination of fabricated micro- trenches in the supporting substrate and focused laser irradiation. This allowed local control of thermally activated oxidation reactions of the nanowire on the scale of the diffraction limit. The locality of the oxidation was detected by micro-Raman mapping, and the results were found consistent with numerical simulations of the temperature profile. Applying the technique to nanowires in electrical devices the locally oxidized nanowires remained conducting with a lower conductance as expected for an effectively thinner conducting core.

Semiconductor nanowires (NWs) offer a range of properties which makes them attractive to diverse areas of research and technology. The reduced size and high surface-to-volume ratio make them promising candidates for advanced electronics[1] and quantum devices[2,3], local chemical and biological sensors[4] as well as efficient solar cells[5] and optical devices[6]. Also, the confinement offered by the NW geometry may allow devices providing greatly enhanced thermoelectric properties[7,8]. In order to exploit the potential of semiconductor NWs for devices with improved or new functionalities, methods of locally controlling the properties along the NWs are often required. Such control has been demonstrated either by advanced device architectures incorporating local electrostatic gates[9,10] or by modifying the crystal properties along the wire by changing material composition during or after growth[11–13]. An alternative approach which has recently emerged is the possibility of using a high intensity focused laser beam to locally induce chemical changes in the NW[14]. By combination of Raman spectroscopy[15] and transmission electron microscopy[16] it has been established that for the case of InAs the high intensity laser, in the presence of ambient air conditions, promotes the oxidation process $As_2O_3 + 2InAs \rightarrow In_2O_3 + 4As$ which converts the irradiated parts of the NW surface to crystalline arsenic and poly-crystalline indium oxide[16]. However, so far, only little is known about the properties of the resulting structures and the spatial resolution that can be achieved has not been analyzed. A similar process occurs for GaAs, and Yazji, S. et al [17] showed that the thermal conductivity of the NW is dramatically decreased upon irradiation, suggesting this method as a way of designing structures with enhanced thermoelectric properties.

Here we show that by suspending NWs over micro-trenches etched into a $SiO_2$ substrate the temperature profile of the NW can be locally engineered. By tuning the laser intensity, only the suspended parts of the wires will reach a temperature sufficient to activate the oxidation process. Using Raman spectroscopy we demonstrate the viability of this approach, achieving a spatial resolution of at least 250 nm; the size of the detection area of our Raman setup. Finally, we study the electrical properties of the oxidized NWs and show that they maintain their ability to carry electrical current and also preserve the key semiconductor property of electrostatic tunability using a nearby gate electrode.

The NWs used for this study were grown by molecular beam epitaxy (MBE) on [111]B InAs substrates covered with gold catalyst particles. The NWs attain the wurtzite crystal structure with the [0001] direction along the NW axis and grow to a length of ~7 μm with a tapered profile having diameters of ~130 nm at the base and 40 nm at the tip[18]. Substrates of degenerately doped silicon with a 500 nm $SiO_2$ capping were prepared with metal alignment grids and arrays of 150 nm deep and 1 μm wide trenches fabricated using electron beam lithography and shallow angle Kaufmann ion milling. The NWs were ultrasonically suspended in isopropanol and randomly dispersed on the substrates. NWs bridging the trenches were located by optical inspection. Figure 1(a) schematically shows the setup and Fig. 1(b) shows an SEM micrograph of a typical sample.

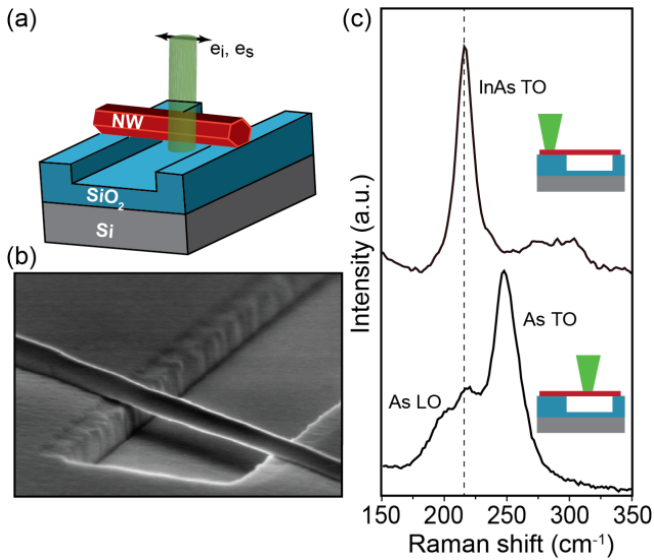

Figure 1. (a) Schematic illustration of the measurement setup and the definition of the coordinate system. $e_i$ and $e_s$ are the polarizations of the incident and scattered laser light, respectively, which are parallel to the NW axes. (b) Tilt view SEM image of an InAs NW with diameter of 100 nm suspended over a 1 μm wide trench. (c) Raman spectra of a suspended InAs NW, recorded as indicated in insert schematics.

The laser irradiation and Raman characterization was carried out using a home built setup consisting of an inverted microscope (Olympus IX71) with a 100x 1.3 NA oil immersion objective (Olympus UPLFLN 100×) and a 70/30 beamsplitter (XF122 from Omega Filters) which focuses a 514.5nm laser beam from an Ar-ion laser (CVI Melles Griot 35MAP431-200) into a ~250 nm diameter spot on the sample. The scattered light was collected through the microscope and analyzed using a PI Acton SpectraPro SP-2356 poly-chromator and a PI Acton SPEC-10:100B/LN_eXcelon spectroscopy system. Sample positioning and scanning with nanometer precision was enabled by a piezo-scanning $xy$ stage (Physik Intrumente P5173CL) on the microscope. The polarization of both incident and scattered light was oriented parallel to the NW axis unless otherwise noted. Electrical access to the contacted NWs on the inverted microscope was achieved using long on-chip, lithographically defined electrodes separating areas for wire-bonding from the microscope objective.

In the following we first focus on samples of suspended NWs without electrical contacts. Figure 1(c) shows typical examples of Raman spectra from a suspended segment and from a spot outside the trench of the same NW with an irradiance of ~450 kW/cm$^2$. The spectrum acquired outside the trench shows the conventional InAs spectrum with the transverse optical (TO) mode appearing at 216 cm$^{-1}$. For the wire part over the trench, in addition to the InAs peak the spectrum contains two modes at ~195 and ~260 cm$^{-1}$. From previous studies combining Raman spectroscopy and TEM analysis these additional modes can be assigned to crystalline arsenic[16] appearing as a result of the thermally activated oxidation process as described above. The results of Fig. 1 thus show that over the trench the irradiation increased the temperature of the NW to have a considerable oxidation while the temperature was not sufficiently high for the NW segment on the substrate to form a detectable amount of arsenic.

To study the spatial variation and locality of the oxidation process, Raman spectra were recorded along the suspended NWs with a step-size of ~100 nm and 60 s integration time in parallel polarization configuration; the arsenic modes do not appear for perpendicular polarization. Figure 2 shows typical results. The particular device is shown in Fig. 2(a) with indications of the positions of the recorded spectra. Figure 2(b) maps the normalized spectra as a function of Raman intensity and scan position. While moving from the supported to the suspended part of the NW, the arsenic spectrum appears abruptly over a distance of ~200 nm, as evident in Fig. 2(c). This scale is comparable to the size of the laser beam, which was used both for inducing and detecting the oxidation reaction.

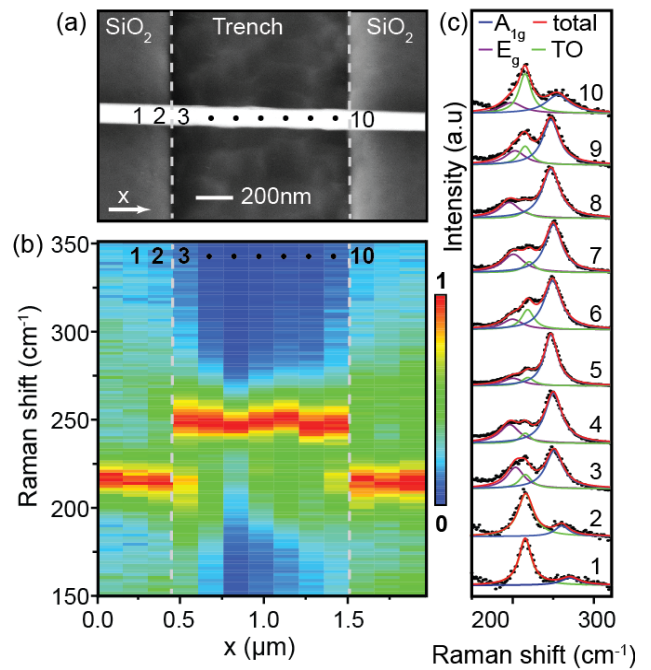

Figure 2. (a) SEM image of the measured NW. The NW has a diameter of ~100 nm and is suspended over a 1 μm trench. The numbers indicate the position along the NW where spectra were recorded. The dashed line in (a) and (b) indicate the position of the trench. (b) Color representation of the Raman intensity as a function of Raman shift and position. Each spectrum was normalized to the maximum As $A_{1g}$ peak intensity. (c) Lorentzian deconvolution of the spectra in (b) with the corresponding numbers.

The oxidation rate grows exponentially with temperature and is expected to be significant for temperatures above 350 °C[19]. The temperature gradient during laser irradiation was investigated by numerical simulations using COMSOL Multiphysics. The model consists of a 1000 µm³ Si block with boundaries fixed at 293 K, a 200 nm thick SiO$_2$ capping layer with a 150 nm deep and 1 µm wide trench bridged by a 7 µm long hexagonal InAs NW with a diameter of 100 nm. Bulk materials parameters were used except for the NW thermal conductivity where the experimental value, $k_{InAs}$ = 10 W/mK, was used[20]. Assuming a Gaussian beam profile, the known reflectivity of InAs at 514.5 nm and total absorption, we find that the NW absorbs ~60 % of the energy of the incoming beam.

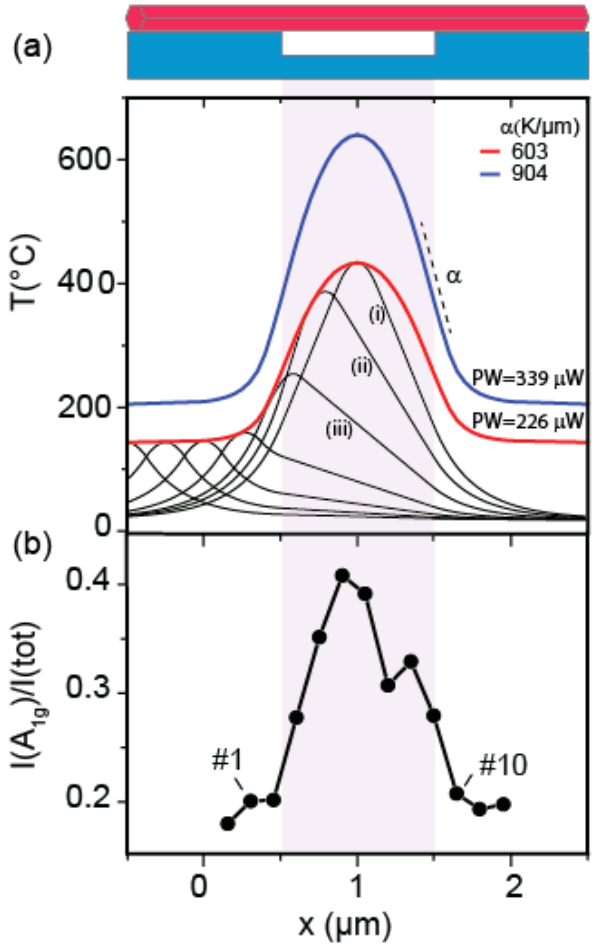

Figure 3. Calculated temperature profile along a 100 nm diameter InAs NW suspended over a 1 µm trench. Top panel shows the geometry. (a) shows the temperature profile along the NW, (i), (ii) and (iii) are for laser positiones at the center of the trench, 0.25 µm from the center and at the edge of the trench, respectively. The red curve shows the maximum temperature attained at position x when scanning the sample with the laser. The blue curve shows the corresponding curve for a 50 % increase in the laser power (PW). The maximum temperature gradient α appears at the edge of the trench. (b) The integrated area of the arsenic $A_{1g}$ mode normalized to the total inegrated area extracted from Fig. 2(b); numbers refer to spectra in Fig. 2.

Figure 3(a) shows the temperature profile (black lines) of the suspended NW, as in the sketch on top, simulated for various positions of the laser using a power chosen to match that of the experiment. The maximum temperature was then found for each data set and plotted as a function of the position along the NW (red trace). Because of the reduced heat dissipation, significantly higher temperatures are reached at the suspended part of the NW and the maximum temperature is achieved when the laser excites the NW at the centre of the trench. Close to the trench edges we find a considerable temperature gradient α of ~600 K/µm resulting in a temperature difference of 120 K over 200 nm. This provides an explanation for the abrupt appearance of the $A_{1g}$ arsenic mode over the trench. Figure 3(b) shows the integrated area of the arsenic $A_{1g}$ normalized to the total integrated area along the NW extracted from spectra in Fig. 2(b,c). The increase follows the calculated temperature with a minor dip in the middle region of the trench.

The results of Fig. 2 and 3 show that engineering the substrate can act as an effective route to rationally determine the positions where oxidation should occur along the NWs. As the oxidation process, leading to the formation of crystalline arsenic, occurs at the surface and the Raman spectra of the oxidized part of the NW in Fig. 2 contain the InAs TO mode, we expect that a significant part of the InAs core remains intact. This has previously been demonstrated by TEM for oxidized InAs NWs suspended on TEM membranes[16]. To investigate the electrical properties of the oxidized NWs, electrical devices consisting of InAs NWs bridging a 1 µm trench and contacted (prior to oxidation) on both sides by 10/120 nm of Ti/Au Ohmic contacts were fabricated (Fig. 4(a)). The highly doped Si back plane of the chip acted as a gate electrode to electrostatically tune the carrier density of the suspended NW when biased by a voltage $V_g$. Figure 4(c) shows the conductance of the NW as a function of $V_g$ before laser irradiation. The conductance is enhanced with increasing $V_g$ showing that the NW acts as an n-type semiconductor which is common for InAs NWs[21] grown without intentional doping. Figure 4(b) shows the resulting Raman spectrum after irradiating the suspended part of the wire with an irradiance of ~450 kW/cm². The arsenic modes characteristic of the oxidation process are clearly observed. We note that relative to the InAs TO mode, the arsenic peak has a lower intensity compared to the sample without electrical contacts in Fig. 2. This is generally observed for contacted NW, and suggests a lower temperature of the contacted NWs due to more efficient thermal anchoring by the metal contacts. This is in agreement with numerical simulations of the contacted nanowire geometry yielding a maximum temperature of the suspended segment of 293 ℃, considerably below the uncontacted case. The electronic conductance of the NW after the oxidation is shown in Fig. 4(d). While the overall conductance has dropped by a factor of 1.5-2 the qualitative dependence remains similar to the pristine NW, and upon applying gate voltages up to 40 V the conductance level is restored. Since in InAs NWs, a significant part of the current may be carried by a surface accumulation layer due to a pinning of the Fermi level in the conduction band at the surface[22,23], we attribute the decreased conductivity to a combination of a thinner NW, increased scattering due to increased surface roughness after the irradiation[16,17] and a higher threshold gate voltage as expected for an effectively thinner InAs core[21].

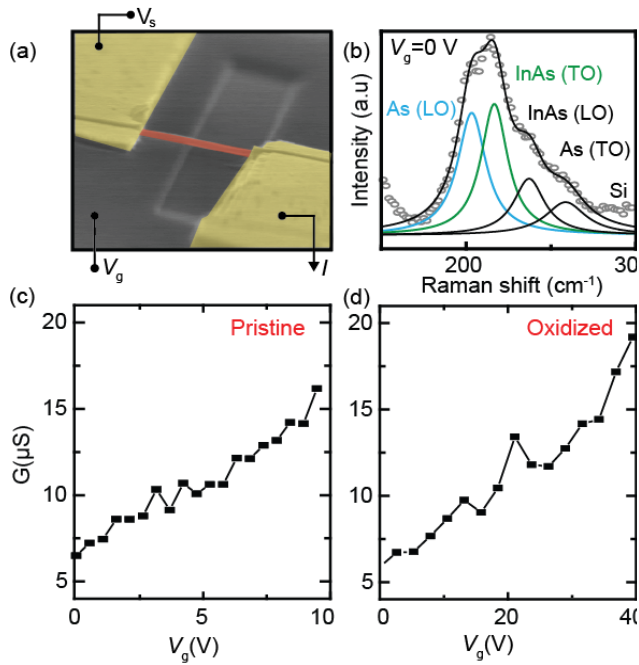

Figure 4. (a) Artificially colored SEM image of the measured device. NW is red, and yellow regions are the Ti/Au contact pads. The width of the trench is 1 μm. (b) Deconvoluted Raman spectra of the NW at zero gate voltage. The peak at 300 cm$^{-1}$ is the silicon peak which was not considered in the fit. (c) and (d) Conductance of the suspended NW as a function of applied back gate voltage before and after laser irradiation.

The observed conservation of the electrical properties of the InAs NW after activating the surface oxidation reaction, opens up for the use of this technique to engineer local functionality to electrical NW devices. As an example, it was shown in Ref[17] that for GaAs NWs the analogous oxidation process, which converts the GaAs to $Ga_2O_3$ and crystalline arsenic, significantly decreases the thermal conductivity of the NWs. If the electrical conductivity remains as shown in Fig. 4(d) this suggest an enhanced thermoelectric effect in the treated NWs which may also be further increased by the large temperature gradients caused by the trench geometry[7]. The reduced diameter of the oxidized NW core can be implemented in electrical devices by inducing barriers in conducting channels[11] or thinning the active regions of field-effect sensors[24]. Furthermore the locality of the laser-induced process which results in left-over metallic arsenic may provide a route for welding of NWs into electrically connected circuits or networks[25], and by performing the irradiation in an controlled environment possibly containing metalorganic gasses the technique may allow control of the induced reaction.

In conclusion, we showed that by a combination of engineered micro trenches in silicon substrates and focused laser irradiation, local oxidation on the scale of ~250 nm could be induced in InAs NWs. The local surface oxidation was confirmed by micro-Raman spectroscopy and was found to be consistent with numerical simulations of the temperature profile of the suspended NW. By incorporating suspended NWs into electrical devices we showed that the electrical properties of the oxidized InAs NWs are consistent with a conducting semiconductor core of reduced diameter. This demonstrates the viability of this technique for engineering functionalities into electrical devices.

## Acknowledgement

We thank the Danish Agency for Science Technology and Innovation (The Danish Council for Strategic Research−ANaCell project), "Center for Synthetic Biology" at Copenhagen University funded by the UNIK research initiative of the Danish Ministry of Science, Technology and Innovation (Grant 09-065274), bioSYNergy, University of Copenhagen's Excellence Programme for Interdisciplinary Research, the Lundbeck Foundation, and the Carlsberg Foundation. The Center for Quantum Devices is supported by the Danish National Research Foundation.